\documentclass[10pt,aps,prl,twocolumn]{revtex4}
\usepackage{graphicx} % Include figure files
\usepackage{subfig}
\usepackage{bm}% bold math
\usepackage{color}
\usepackage[english]{babel}
\usepackage{amsfonts,amssymb,amsmath}
\usepackage[normalem]{ulem}
\renewcommand{\v}[1]{\ensuremath{\mathbf{#1}}} % for vectors
 
\newcommand{\abs}[1]{\left| #1 \right|} % for absolute value

\begin{document}

\title{A correction for the Hartree-Fock Density of States for Jellium without Screening} 
 
\author{Alexander I. Blair, Aristeidis Kroukis and Nikitas I. Gidopoulos}
\affiliation{Department of Physics, Durham University, South Road, Durham, DH1 3LE, United Kingdom }

\begin{abstract}

We revisit the Hartree-Fock (HF) calculation for the uniform electron gas, or jellium model, whose predictions -- divergent 
derivative of the energy dispersion relation and vanishing density of states (DOS) at the Fermi level -- 
are in qualitative disagreement with experimental evidence for simple metals.
Currently, this qualitative failure is attributed to the lack of screening in the HF equations.

Employing Slater's hyper-Hartree-Fock (HHF) equations, derived variationally, to study the ground state and the excited states of jellium, we find that the divergent derivative of the energy dispersion relation and the 
zero in the DOS are still present, but shifted from the Fermi wavevector and energy of jellium 
to the boundary between the set of variationally optimised and unoptimised HHF orbitals. 
The location of this boundary is not fixed, but it can be chosen to lie at arbitrarily high values of wavevector and energy, 
well clear from the Fermi level of jellium.

We conclude that, rather than the lack of screening in the HF equations, 
the well-known qualitative failure of the ground-state HF approximation
is an artifact of its nonlocal exchange operator.
Other similar artifacts of the HF nonlocal exchange operator, not associated with the lack of electronic correlation, are known in the literature.

\end{abstract}


\maketitle

\section{Introduction} 

The uniform electron gas, or jellium model, is an archetypal example in solid-state physics and many-body theory.
Its treatment, in the Hartree Fock (HF) approximation, can be found in classic textbooks \cite{thouless,ashcroftmermin,kittel,fulde,inkson,lb}, 
where, we learn that the 
HF equations applied to the ground state of the jellium, 
admit plane wave solutions with energy-wavevector dispersion relation given by, 
\begin{align}
\label{dispHF}
\varepsilon(k) = \frac{k^2}{2} - \frac{k_F}{\pi} \left(1 + \frac{k_F^2 - k^2 }{2 k k_F} \ln\abs{\frac{k_F + k}{ k_F - k}}\right)\, .
\end{align} 
%where $k\equiv \abs{\v k}$, and 
$k_F$ is the Fermi wavevector,
%\begin{equation}
$k_F^3 = 3 \pi^2 { ( N / V ) } $.
%\end{equation}
%
The single-particle energy $\varepsilon (k) $ %in (\ref{dispHF}) 
is the sum of the free-electron energy, $k^2 / 2$, and the single-particle 
exchange energy. 
The Fermi wavevector $k_F$ is often expressed in terms of the mean radius per particle $r_s = \sqrt[\leftroot{-1}\uproot{2}3]{9\pi/4 k_F^3}$ \cite{fulde}; for 
typical values of $r_s$ in metals, the two terms in (\ref{dispHF}) are comparable in size.

\begin{figure}[h]
\begin{center}
%\centering
\includegraphics[trim =7mm 7mm 5mm 5mm, clip, width= 6cm, height =4cm ]{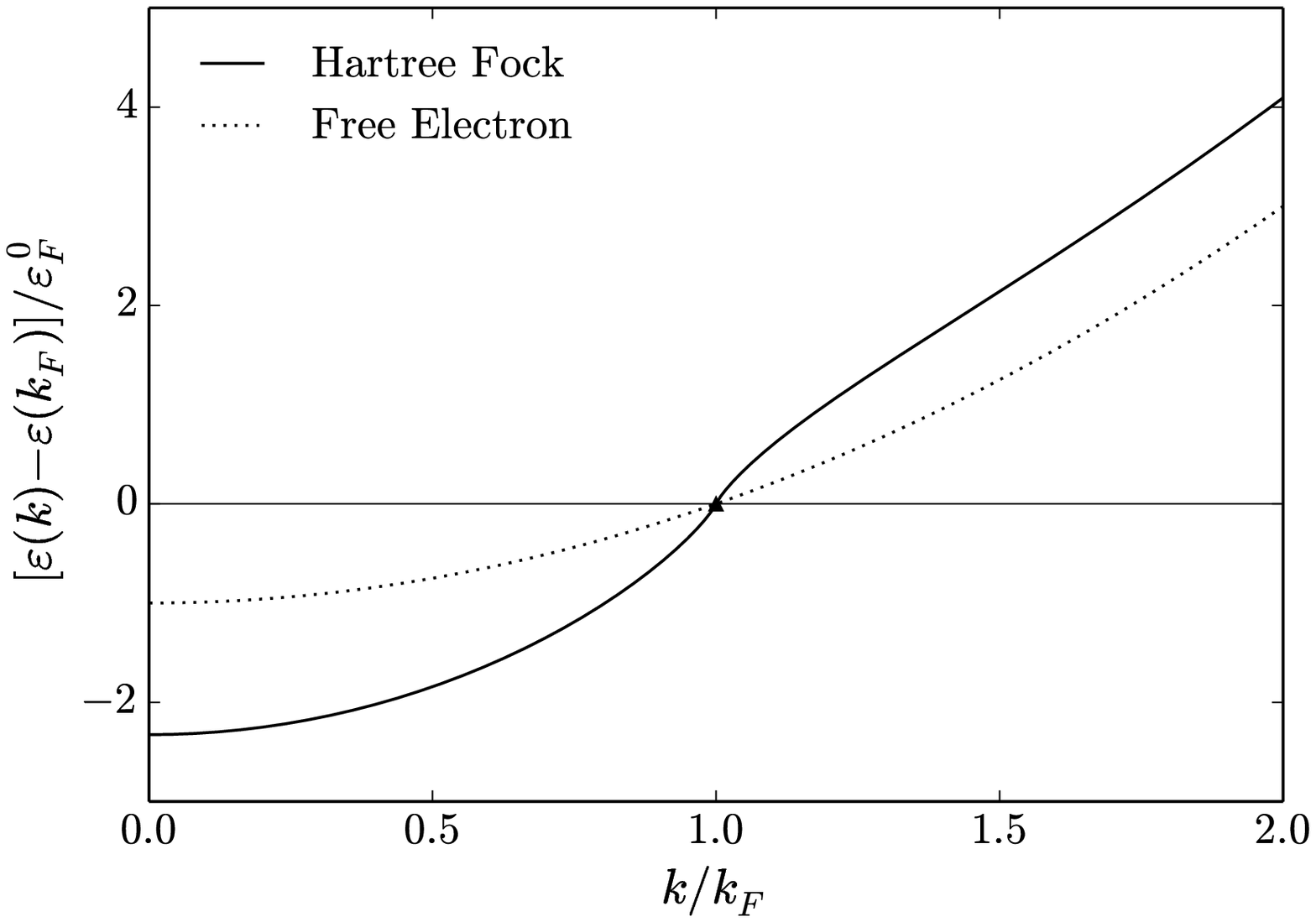} 
\includegraphics[trim =7mm 7mm 5mm 5mm, clip, width= 6cm, height =4cm]{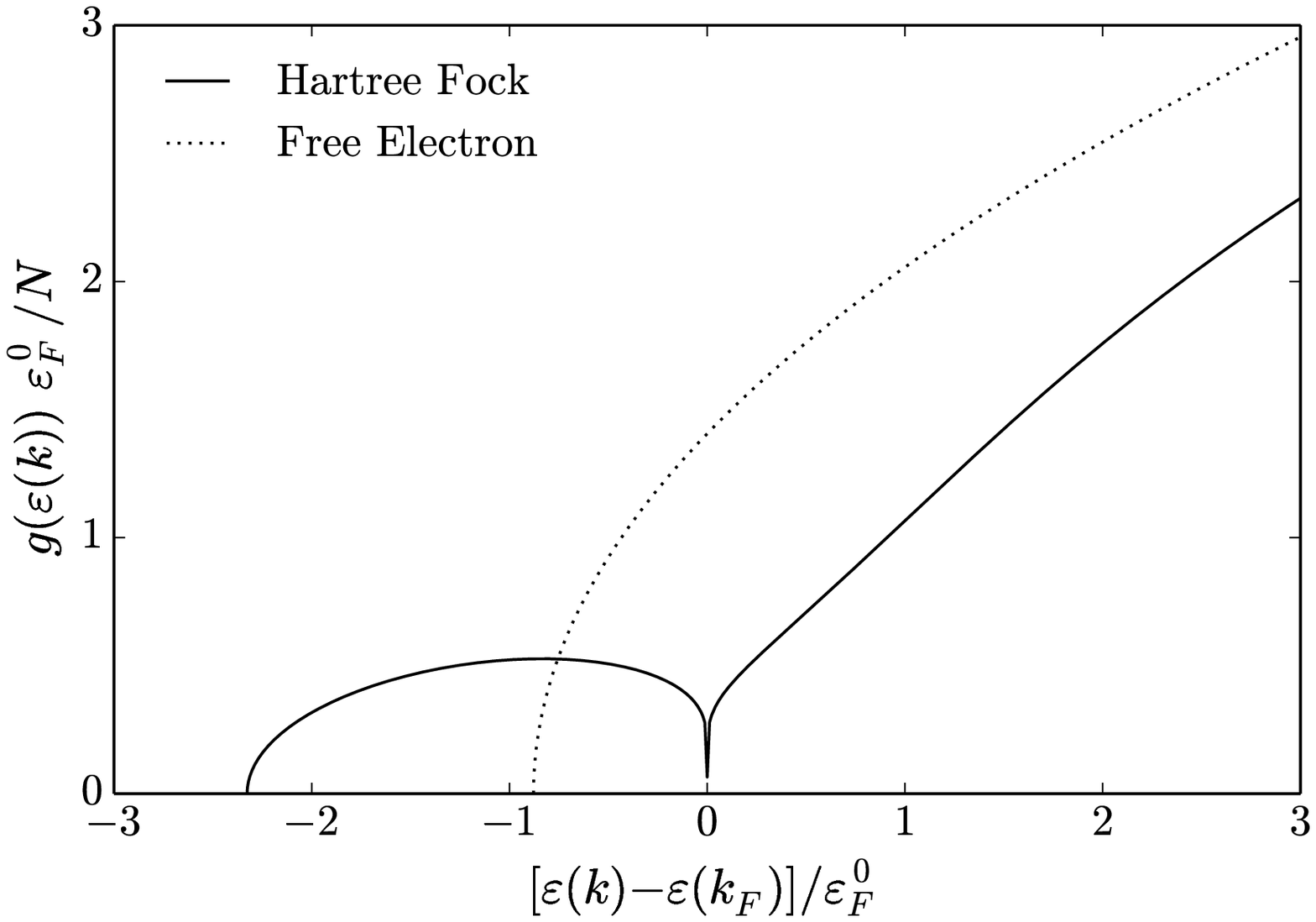} 
\end{center}
\caption[]{Solid lines show ground-state HF results for jellium, compared to free-electron results in dotted lines. ($r_s/a_0 =4$, $\varepsilon^0_F = k^2_F/2$.)
Top: Energy vs wave vector dispersion relation $\varepsilon (k)$. % for jellium ($r_s/a_0 =4$), compared to free-electron result. 
The logarithmic divergence in the derivative, $d \varepsilon / d k$, is marked with a triangle ($\blacktriangle$). % marks the logarithmic divergence of the derivative of the HF dispersion at the Fermi level. 
Bottom: %Ground-state HF 
DOS, showing the unphysical zero at the Fermi level for jellium.} %($r_s/a_0 =4$) 
%shows unphysical zero at the Fermi level.}%, compared to free electron result.}
%$\varepsilon$:~independent-particle level; $k_{(F)}$:~magnitude of (Fermi) wave vector; $r_s$:~mean radius per particle;~$a_0$: Bohr radius.}
\label{fig:HF}
\end{figure}

It is well known in the literature that the 
dispersion relation (\ref{dispHF}) has a logarithmically divergent derivative at the Fermi energy, shown in Fig.~\ref{fig:HF}.
% in contrast to the free electron result.
Another marked difference between the free electron result and the HF solution for jellium, evident in Fig.~\ref{fig:HF}, 
is the considerably increased bandwidth of the HF dispersion.
Finally, it is well known that in the HF approximation the DOS for jellium vanishes at the Fermi level (Fig.~\ref{fig:HF}), 
since the DOS is inversely proportional to the derivative of the dispersion. The zero in the
DOS at the Fermi level suggests that jellium is a semimetal, in obvious disagreement with experimental evidence for simple metals, 
such at sodium or aluminium, which are described accurately by the jellium model.

In the literature, the qualitatively wrong description of jellium in the HF approximation is attributed to the long range of the Coulomb repulsion 
\cite{thouless,ashcroftmermin,kittel,fulde,inkson,lb}.
It is well known that the flawed description can be corrected by the introduction of electronic many body correlation 
effects \cite{thouless,ashcroftmermin,kittel,fulde,inkson,lb,PhysRevA.36.888}, 
which screen the bare Coulomb potential and thus eliminate the unphysical divergent derivative of the dispersion relation, the zero in the DOS at the Fermi level, and also 
reduce the bandwidth of the HF dispersion relation of jellium.

In an effort to understand whether HF's lack of screening actually plays a role, we revisit the HF study of jellium, attempting to correct the qualitative errors of the HF description, but without including any form of electronic correlation. 
For this purpose, we employ Slater's hyper-HF (HHF) theory for the ground and the excited states of an $N$-electron system \cite{hhf}.
Specifically, we use the single-particle HHF equations by Gidopoulos and Theophilou \cite{gt,footnote}, 
who considered an $N$-electron system described by a Hamiltonian $H$ and then variationally 
optimised the average energy $\sum_n \langle \Phi_n | H | \Phi_n \rangle $ of all configurations 
($N$-electron Slater determinants $\Phi_n$) constructed from a basis set of $R$ spin-orbitals, $R \ge N$.

\section{The hyper-Hartree-Fock equations for jellium}

The aim in HHF theory is to obtain approximations, at the HF level of description, for the ground and the excited 
states of an $N$-electron system. These states are represented by $N$-electron
Slater determinants, constructed from a common set of spin-orbitals. 
Obviously, to have the flexibility to describe excited states, the number of spin-orbitals
($R$) must exceed the number of electrons. 
For example, say we are interested to approximate the ground and excited states of the helium atom.
In the HF ground state of He, the $1s$ orbital ($\varphi_{1s}$) is doubly occupied.   To study a couple of  
excited states, we need at least one more spin-orbital and the next one is: $\varphi_{2s}^{\uparrow}$.
With the three available spin-orbitals, $1s^{\uparrow},1s^{\downarrow},2s^{\uparrow}$, we can form three configurations for the He atom (two-electron Slater 
determinants): $\Phi_1 = [1s^{\downarrow},1s^{\uparrow}]$, $\Phi_2 = [1s^{\downarrow}, 2s^{\uparrow}]$, $\Phi_3 = [1s^{\uparrow}, 2s^{\uparrow}]$. In HHF theory, we variationally optimise the three common 
spin-orbitals simultaneously, by minimising the sum of the expectation values 
$\sum_{i=1}^3 \langle \Phi_i | H | \Phi_i \rangle$.
The minimisation leads to the HHF single-particle equations for the three spin-orbitals. 
It turns out that these equations resemble the ground-state HF equations for the lithium atom (three electrons) 
but with a weakened Coulomb repulsion between the three electrons, to keep balance with the nuclear 
charge which has remained that of the He nucleus.

In general, in HHF theory \cite{hhf,gt} for an $N$-electron system, one considers a set of $R$ orthonormal spin-orbitals, with $R \ge N$.
%, where $N$ is the number of electrons of the system of interest. 
On this {\color{black} spin-orbital} basis set one may define, $ D = R!/(N!(R-N)!)$, $N$-electron Slater determinants.

The derivation of the single-particle HHF equations in Ref.~\onlinecite{gt} is based on Theophilou's variational principle \cite{theophilou1979energy},
\begin{equation}
\label{Theo}
\sum^D_{n=1} \langle \Phi_n|{H}| \Phi_n\rangle \geq \sum^D_{n=1} E_n^{(0)} \, ,
\end{equation}
where $\{ E_n^{0} \}$ are the $D$ lowest eigenvalues of the $N$-electron Hamiltonian $H$. 

An extension of the variational principle, with unequal weights in the sums in (\ref{Theo}) was proposed by Theophilou  \cite{theoGOK}, and independently by 
Gross, Oliveira, Kohn \cite{PhysRevA.37.2805}. These variational principles can be derived as special cases from the Helmholtz variational principle in statistical mechanics \cite{rajagopal,PhysRevA.87.062501}. 
In particular, the inequality in (\ref{Theo}) arises as the high temperature limit of the Helmholtz
variational principle.

Optimisation of the $R$ spin-orbitals $\{ \varphi_i \}$ to minimise 
the sum of the energies on the l.h.s. of (\ref{Theo}) leads to the following single-particle equations 
for the spatial part of the spin-orbitals \cite{gt} (in atomic units): %. Optimising a single spin-orbital $\varphi_i$, we arrive at the equation 
\begin{align}
\label{subHF}
&\left[ -\frac{1}{2}\nabla^2 + V_{\text{ext}}(\v r) \right] \varphi_i(\v r) \nonumber \\
&+ {1 \over \Lambda} \, \sum^R_{j=1}\left[J_j(\v r) - \delta_{s_j,s_i} K_j(\v r) \right]\varphi_i(\v r)  = \lambda_i \varphi_i(\v r) \, , \end{align}
where, 
\begin{equation} \label{L}
\Lambda = {R-1  \over N-1} \, , 
\end{equation}
and
\begin{align}
J_j(\v r) \varphi_i(\v r)  &\equiv \int\frac{ d^3 \v r'}{ \abs{ \v r - \v r'}} \abs{\varphi_j(\v r')}^2 \varphi_i(\v r) \, ,\\
K_j(\v r) \varphi_i(\v r)  &\equiv \int\frac{ d^3 \v r'}{ \abs{ \v r - \v r'}} \varphi^*_j(\v r') \varphi_i(\v r') \varphi_j(\v r)\, ,
\end{align}
are the Coulomb and exchange operators respectively. 
%Inner products in spin space have been taken for the Coulomb and exchange operators. 
$V_{\text{ext}}$ signifies the attractive potential of the nuclear charge. 
%, and the eigenvalue $\lambda_i$ is chosen such that the coefficient multiplying the single-particle kinetic energy operator is unity. 
For $R=N$, Eqs.~\ref{subHF} reduce to the familiar ground-state HF equations. 
In Eqs.~\ref{subHF}, the orbitals $\varphi_i$, with $i = 1 , \ldots , R$, are correctly repelled electrostatically by a charge of $N-1$ electrons.
In contrast to the ground-state HF case \cite{RevModPhys.44.451}, 
this holds true even for the orbitals that are not occupied in the {\color{black} HHF} ground-state Slater determinant
as long as these orbitals are variationally optimised, 
i.e., for $\varphi_i$, with $N < i \le R$. 
Furthermore, the orbitals that are left variationally unoptimised, i.e., $\varphi_i$, with $i > R$, are repelled by a charge
of $N-1 + (1/\Lambda)$ electrons. 
In the HHF equations, the well-known asymmetry in the treatment of the
variationally optimised and unoptimised orbitals by the nonlocal exchange operator \cite{RevModPhys.44.451} is 
still present, but softened (for large $\Lambda$), compared with ground-state HF. 
We note that for $R > N$, Koopmans' theorem \cite{koopmans,szabo} ceases to hold for the HHF equations.

The HHF equations (\ref{subHF}) have the form of ground-state HF equations for a virtual system of $R$-electrons, where the electronic 
Coulomb repulsion is multiplied by $1 / \Lambda$: $1/|{\bf r} - {\bf r}'| \rightarrow \Lambda^{-1}/|{\bf r} - {\bf r}'|$.  
Therefore, the calculation of the optimal spin-orbitals to represent the ground and 
excited-states of an $N$-electron system in the HHF approximation, is reduced to the 
calculation of the ground-state HF orbitals of a fictitious system with a greater number of electrons $R \ge N$, and scaled down 
electronic Coulomb repulsion. % or enhanced electron-nuclear Coumb attraction.
A related approach is the ``super-hamiltonian method'' by Katriel \cite{k1,k2}.

\color{black}
Finally, before applying the HHF equations to jellium, 
we remark that correlated, approximate eigenstates of the Hamiltonian 
$H$ can be obtained by diagonalising the matrix $\langle \Phi_n | H | \Phi_m \rangle$~\cite{gt}, 
where $\Phi_n$ are the $N$-electron HHF Slater determinants.
This configuration-interaction method employs the HHF spin-orbitals, %in a subspace of the Hilbert space 
which are optimised to represent on equal footing the ground and the excited states of $H$. 
%It is interesting to carry out such CI or other correlated, post-HF calculations based on the HHF spin-orbitals
%and compare them with similar calculations based traditionally on the ground-state HF spin-orbitals.

\subsection{Solution of the HHF equations for jellium.}

\color{black}
Similarly to the HF ground state, the HHF equations for jellium admit plane wave solutions. 
\color{black}
It follows that the ground-state $N$-electron Slater determinant and its total energy are 
the same in the HF and HHF approximations.   

Although the HF and HHF equations for jellium admit the same solution for the orbitals, 
the dispersion relations for the single-particle energies $\varepsilon (k)$ and $\lambda (k) $ differ.
In particular, the HHF dispersion, $\lambda (k)$, 
results from an optimisation that involves a broader range of wavevectors than the HF dispersion $\varepsilon (k)$.   
\color{black}

\begin{figure}[h]
\begin{center}
\includegraphics[trim =6mm 6mm 5mm 4mm, clip, width= 8cm, height =6cm ]{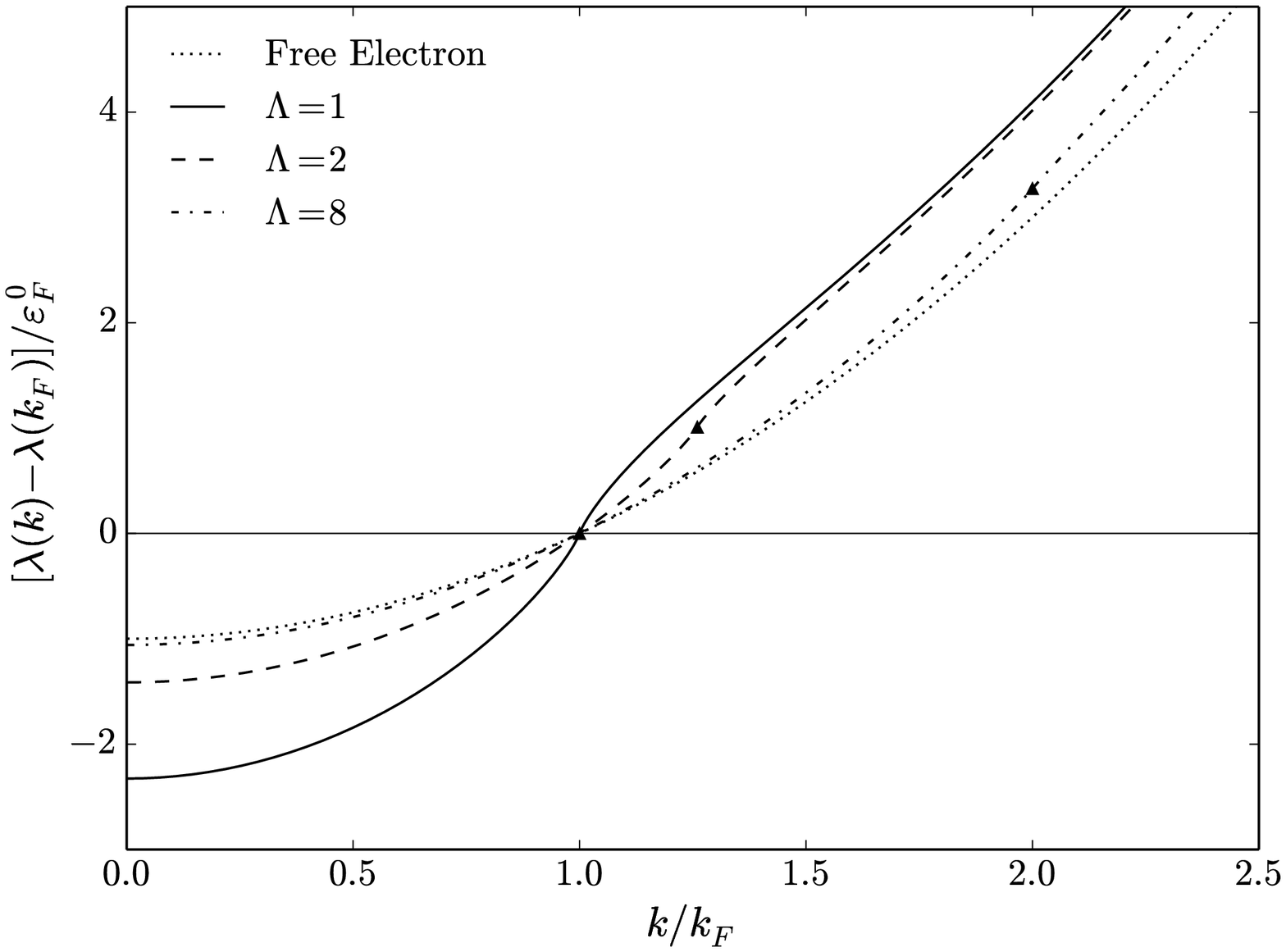} 
\includegraphics[trim =6mm 6mm 5mm 4mm, clip, width= 8cm, height =6cm ]{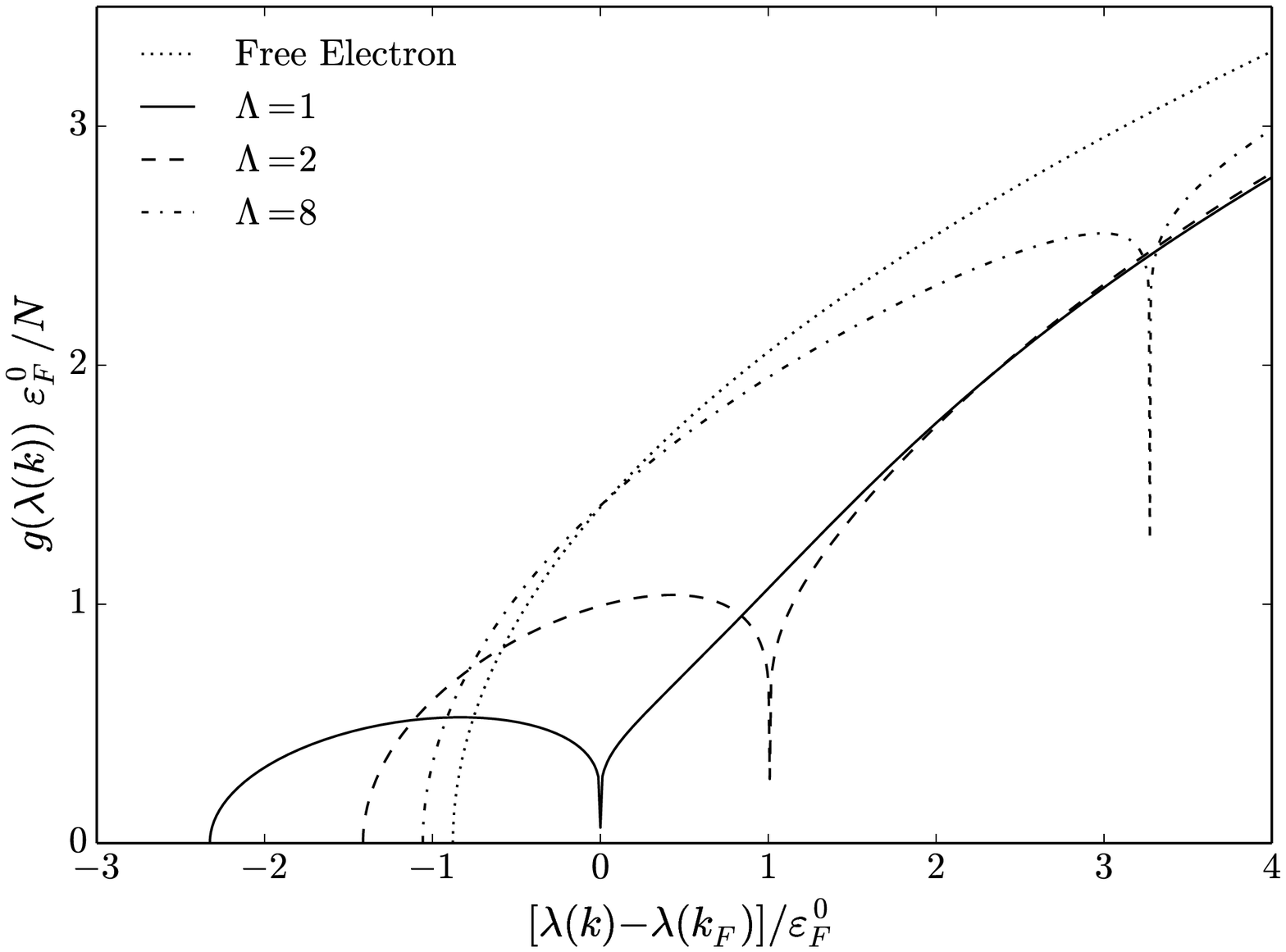}  
\end{center}
\caption[]{Excited-state HF results for jellium, for various $\Lambda = R/N$, compared to free electron results (dotted lines). 
($r_s/a_0 =4$, $\varepsilon^0_F = k^2_F/2$.)
Top: Energy vs wave vector dispersion relations $\lambda (k)$. % for jellium ($r_s~=~4~a_0$), for various $\Lambda = R/N$, compared to free electron result. 
When $\Lambda = 1$, $\lambda(k) = \varepsilon(k)$.
Triangles ($\blacktriangle$) mark logarithmic divergence in $d \lambda/ d k$ at Fermi level of fictitious $R$-electron system. 
Bottom: DOS $g(\lambda(k))$, showing the zero at the Fermi level of the fictitious $R$-electron system. 
}
\label{fig:SS}
\end{figure}

Following the standard treatment in textbooks \cite{thouless,ashcroftmermin,kittel,inkson,fulde,lb}, it is straightforward to work out 
directly the solution of the HHF Eqs.~\ref{subHF}.
Here, we exploit the similarity of Eqs. \ref{subHF} with ground-state HF equations of an $R$-electron system, 
%with scaled down electronic Coulomb repulsion. Consequently, 
to obtain that the HHF dispersion relation, $\lambda (k) $, will be given by (\ref{dispHF}) with the 
single-particle exchange energy scaled down by the factor $1/ \Lambda$:

\begin{align}
\label{krdisprel0}
\lambda(k) = \frac{k^2}{2} 
- \frac{k_R}{\Lambda \pi } \left(1 + \frac{k_R^2 - k^2 }{2 k k_R} \ln\abs{\frac{k_R + k}{k_R - k}}\right) \, .
\end{align}
$k_R$ is the Fermi wave vector of the virtual $R$-electron system,
\begin{equation}
k_R^3 = 3 \pi^2 \, {R \over V} \, .
\end{equation}
Dividing $k_R / k_F$, and taking the thermodynamic limit, $N, V \rightarrow \infty$, with the ratio $\Lambda$ fixed, we obtain:
\begin{equation}
 k_R = \Lambda^{1/3} k_F \, .
\end{equation}
Substitution of the above into Eq. (\ref{krdisprel0}) yields the desired expression for the single particle energy levels of jellium, 
in terms of $\Lambda$ and the Fermi wavevector $k_F$ of the actual $N$-electron system:
\begin{align}
\label{disprelation0}
\lambda(k) =
 \frac{k^2}{2} -  & \Lambda^{-2/3}  \frac{k_F}{ \pi}    \\ \nonumber
 & \times \Biggl(1+ \frac{\Lambda^{2/3} k_F^2 - k^2 }{2 \Lambda^{1/3} k k_F} \ln\abs{\frac{\Lambda^{1/3}k_F + k}{\Lambda^{1/3} k_F - k}}\Biggr) \, .
\end{align} 
The dispersion relation in Eq. (\ref{disprelation0}) reduces to the ground-state 
Hartree-Fock result for $\Lambda = 1$, and to the free electron dispersion relation, 
$\lambda (k) = k^2 / 2$, in the limit $\Lambda \rightarrow \infty$ (See Fig.~\ref{fig:SS}). 
For increasing $\Lambda$, the bandwidth of the HHF dispersion, $\lambda ( k ) $, decreases compared to the ground-state HF dispersion, 
$\varepsilon ( k )$. For $\Lambda \rightarrow \infty$ the exchange term in HHF dispersion vanishes and $\lambda (k) $ reduces to the free-electron result.

Importantly, the wave vector at which the logarithmically divergent derivative occurs is shifted from $k_F$ to $k_R$, such that the divergence 
no longer occurs at the Fermi energy of the physical $N$-electron system, when the number of optimised orbitals is $R>N$.

The DOS, $g(\lambda) \delta \lambda$, can be obtained directly from Eq.~\ref{disprelation0} \cite{ashcroftmermin} and is given 
by the parametric equation:
\begin{eqnarray}
\lefteqn{g(\lambda(k)) = \frac{Vk^2}{\pi^2 \, (d \lambda / d k )} }\nonumber \\
&=&   \frac{Vk^2}{\pi^2}  \, \Biggl[k - \frac{1}{\Lambda \pi } \left(\frac{k_R}{k} - \frac{ k_R^2 + k^2 }{2  k^2} \ln\abs{\frac{k_R + k}{k_R - k}}\right) \Biggr]^{-1}\, . 
\label{dos}
\end{eqnarray} 
$g(\lambda(k))$ is expressed in terms of $k_R$ (rather than $\Lambda^{1/3} k_F$) to keep the notation simple. 
For any finite $\Lambda$, the DOS still vanishes. However, as shown in Fig. \ref{fig:SS}, the zero in the DOS occurs at the Fermi energy 
of the fictitious $R$-electron system, $\lambda(k_R)$,  rather than the Fermi energy of the physical system $\lambda(k_F)$.

%\vspace{-3ex}
\section{Discussion}
%\vspace{-3ex}

In metals, screening is an important effect that reduces the range of the effective repulsion between electrons, shielding any charge 
at distances greater than a characteristic screening length.
In the literature of many-body theory \cite{thouless,kittel,inkson,fulde,lb} and solid-state physics \cite{ashcroftmermin,kittel,inkson}, 
where jellium is a paradigm, the qualitatively flawed description of metals by 
the HF approximation is attributed to the long-range nature of the Coulomb interaction, which, combined with the 
neglect of correlation, deprives from the HF equations the flexibility to model the phenomenon of screening. 
This understanding of HF's failure is further supported by the softening of the divergence in the slope of $\varepsilon (k) $, after replacing 
the bare Coulomb potential 
in the HF nonlocal exchange term by a screened Coulomb potential \cite{ashcroftmermin}. 
%Interestingly, the preferred derivation of the HF equations in this part of the community is usually the equation of motion
%method, rather than the variational method followed in this work. 

On the other hand, in the theoretical chemistry literature, 
it is well known that the HF nonlocal exchange term, in finite systems, 
gives rise to several counterintuitive results, reminiscent of the HF anomalies in jellium, 
which are not associated with HF's lack of correlation. For example, 
%On the other hand, a plethora of reports have been published on counterintuitive effects of the HF nonlocal exchange operator, on finite systems. 
Handy et. al \cite{PhysRev.180.45} disproved the widely held view \cite{thouless}, 
that the HF nonlocal exchange potential decays as $(-1/r)$ at large distances. 
In particular, in Ref.~\onlinecite{PhysRev.180.45} it was demonstrated that the asymptotic decay of an occupied HF orbital $\varphi_i$
with eigenvalue $\varepsilon_i$ is not $\sim \exp(-\sqrt{-2 \varepsilon_i} r )$, as would be expected from the $(-1/r)$ tail of the exchange potential. 
On the contrary, in HF all the occupied orbitals $\varphi_i$ decay uniformly at large distances away from the system, regardless of their energy eigenvalue. 

%\newpage

As already mentioned, it is also widely known that the HF exchange operator deals with the occupied %(variationally optimised) 
and unoccupied %(variationally unoptimised) 
orbitals in the ground-state HF Slater determinant in an asymmetric way~\cite{RevModPhys.44.451}: 
for an $N$-electron system, the (variationally optimised) 
occupied orbitals are repelled electrostatically by a charge of $N-1$ electrons, while the (variationally unoptimised) 
unoccupied orbitals feel the stronger repulsion of $N$ electrons, making the unoccupied orbitals too diffuse, and raising their energy eigenvalue to unphysically 
high values \cite{RevModPhys.44.451}. 
%
%

%\color{black}

It follows that the energy %$( \langle \Phi_i^a | H | \Phi_i^a \rangle - \langle \Phi | H | \Phi \rangle$) 
to excite an electron from an occupied HF orbital 
$\varphi_i$ to a virtual HF orbital $\varphi_a$, keeping the other occupied orbitals frozen, 
must be smaller than the eigenvalue difference $\varepsilon_a - \varepsilon_i$.
This is because the 
energy $\varepsilon_a$ incorporates the Coulomb interaction of the 
orbital $\varphi_a$, hosting the electron after excitation, with all the occupied
orbitals, including the orbital $\varphi_i$ accommodating the same electron before excitation.
In this sense, 
the Coulomb interaction of the orbitals $\varphi_i$ and $\varphi_a$ 
can be interpreted as a form of self-interaction raising the energy of 
the virtual level $\varepsilon_a$.
It is similar to the ``ghost'' self-interaction discussed in Ref.~\onlinecite{ghost}.
This interpretation is consistent with viewing the virtual HF energies as single-particle levels of the 
$N$-electron system.
%,
Alternatively, by extending the proof of Koopmans' theorem \cite{koopmans}, % to the virtual HF orbitals, 
it can be shown that $\varepsilon_a $ %the eigenvalue of a virtual orbital 
is equal to the negative of the electron affinity of the system to bind an electron at the virtual orbital $\varphi_a$,  
see e.g. Ref.~\onlinecite{szabo}.  
The interpretation of the virtual energies as negative affinities amounts to 
regarding the unoccupied levels as virtual levels of an $(N+1)$-electron system.

We note that for jellium, where the occupied and virtual orbitals are plane waves, 
the self-interaction error of the virtual energy levels (discussed above) 
vanishes in the thermodynamic limit. 
Therefore, it makes sense to consider that the HF 
unoccupied levels represent virtual single-particle levels of the $N$-electron system, 
and to study their dispersion relation and density of states.

%\color{black}
A consequence of the asymmetry in the treatment of occupied versus unoccupied 
orbitals in the HF equations,  %self-interaction error 
is presented by Bach et al. \cite{PhysRevLett.72.2981}, who prove that, for a finite system,
the highest occupied spin-orbital in the ground-state unrestricted-HF Slater determinant, is nondegenerate: 
a nonzero gap separates it 
from the lowest unoccupied spin-orbital, even in systems with an odd number of electrons, in contrast to physical expectation.

Recently, Hollins et al., using two different methods, 
the optimised effective potential, or exact exchange method \cite{us}, 
and the local Fock exchange potential (LFX) method \cite{lfx},  
showed that it is possible to omit correlation and still obtain an accurate dispersion 
relation $\varepsilon (k) $ for simple metals, provided the exchange potential term is local 
${\hat v}_{x} = v_{x} ({\bf r})$ \cite{lfx}. 
%
%In this sense, the recent result of the local Fock exchange (LFX) potential by Hollins et al. \cite{lfx}  is particularly interesting.
The LFX potential, defined as the local exchange potential with the HF ground-state density \cite{rya,lfx}, is particularly interesting in our context: 
even though the determination of the LFX potential is based on the same ground-state HF calculation that gives
a very poor prediction for the dispersion of simple metals (e.g. Na, Al), the band-structure of the LFX potential \cite{lfx} 
almost coincides with the band-structure of the local density approximation in density functional theory \cite{kohanoff2003density}, 
which, by construction, is very accurate for these systems.  
With regard to the vanishing of the HF DOS of metals at the Fermi energy, 
Hollins et al. \cite{lfx} argue that it appears to be an artifact of the HF nonlocal exchange operator.

Our work supports this point of view, by studying the nonlocal HF exchange term directly.
We find that the well-known anomalies of the HF description of jellium are still present in the solution of the 
HHF equations. However, the location of the divergent derivative of $\lambda (k)$ and the zero 
in the DOS are no longer at the Fermi level of the actual $N$-electron system. Instead, they are positioned  
at the border separating the variationally optimised and unoptimised orbitals.
By choosing to variationally optimise a very large number of orbitals, the unphysical zero in the DOS 
can be pushed to very high energies, avoiding completely the window of single-partice energies 
that can be of any relevance to the ground state of jellium.         
Therefore, it is no longer justified to relate the mobile divergence in the derivative of the HHF dispersion and the 
travelling zero in the HHF DOS of jellium, 
with the  lack of electronic correlation in the HHF approximation,
even though these anomalies can still be removed by screening the 
Coulomb repulsion in the exchange term.

This strengthens the view that the failure of the ground-state HF solution for jellium is also an artifact of the nonlocal exchange operator. 
At least part of the explanation of the divergent derivative of the dispersion 
seems to be the 
asymmetry in the treatment of the variationally 
optimised and unoptimised orbitals: as the wavevector $k$ crosses $k_R$ from below, the plane wave solutions 
of the excited-state HHF equations 
are subjected to a discontinuous drop of the nonlocal exchange term and a correspondingly discontinuous 
increase in the Coulomb repulsion from the Hartree term. 
This raises the single-particle eigenvalue $\lambda (k)$ to higher energies, diminishing 
the DOS in the neighborhood of $\lambda (k_R)$.  
The same mechanism operates in the divergent slope of the ground-state HF dispersion $\varepsilon (k)$.

In conclusion, we suggest that the {\color{black} qualitative} failure of the HF approximation for jellium is unrelated to the lack of 
correlation in the HF approximation. 
Instead, this failure is another example in the list of counterintuitive results 
%(or even artifacts), 
caused by the nonlocality of the HF exchange operator.
Our work complements the work of Hollins et al. in Ref.~\onlinecite{lfx} and our conclusion is contrary to what is currently 
written in almost any textbook in the fields of 
many body theory and solid state physics.
To the best of our knowledge, Slater gave the only hint so far in the literature
that the HF failure may be unrelated to electronic 
correlation. In his textbook on ``Insulators, Semiconductors and Metals''  
\cite{hint} he writes that 
``it cannot be in accordance with experiment to write the total energy of the electronically excited state of the crystal as a sum of one-electron energies 
of the type of ...'' (cf. the HF solutions). ``A great deal of effort has gone into explaining this apparent paradox connected with the free-electron theory of electrons in metals. ... It is the impression of the present author, however, that we do not need to look for any deep and profound explanation''.


\begin{thebibliography}{99}


\bibitem{thouless}
D.J. Thouless {\em The Quantum Mechanics of Many-Body Systems} Second Edition, {\em Pure and Applied Physics Series}, Academic Press Inc., New York (1972)

\bibitem{ashcroftmermin}
N.W. Ashcroft, and N.D. Mermin, {\em Solid State Physics}, Saunders College Publishing, Harcourt, Inc., Orlando (1976)

\bibitem{kittel}
C. Kittel, {\em Quantum Theory of Solids}, John Wiley \& Sons, Inc., New York (1976)


\bibitem{inkson}
J.C. Inkson {\em Many-Body Theory of Solids An Introduction} Plenum Press, New York (1984)


\bibitem{fulde}
P. Fulde, {\em Electron Correlations in Molecules and Solids}, Springer, Berlin, Heidelberg (1995)

\bibitem{lb}
T. Lancaster, S.J. Blundell {\em Quantum field theory for the gifted amateur}, Clarendon Press, Oxford ({\color{black} 2014})



\bibitem{PhysRevA.36.888}
%  title = {Fermion gas with screened interactions},
M.A. Ort{\ifmmode \breve{}\else \u{i}\fi}z, and R.M. M\'endez-Moreno, 
Phys. Rev. A, {\bf 36}, 888, (1987); doi:10.1103/PhysRevA.36.888

\bibitem{hhf}
J.C. Slater {\em The Self-consistent Field for Molecules and Solids}, Vol. 4 of the series {\em Quantum Theory of Molecules and Solids}, 
McGraw-Hill Book Inc. (1974) 



\bibitem{gt} N. Gidopoulos and A. Theophilou, Phil. Mag. Part B, {\bf 69}, 1067-1074, (1994); doi:10.1080/01418639408240176.


\bibitem{footnote} 
At the time of publication of Ref.~\onlinecite{gt}, its authors were not aware of Slater's HHF scheme.



\bibitem{theophilou1979energy}
A.K. Theophilou,
%  title={The energy density functional formalism for excited-states},
J. Phys. C: Solid State Physics, {\bf 12}, 
%  number={24},
5419, (1979)






\bibitem{theoGOK}
A.K. Theophilou, in {\em The Single-Particle Density in Physics and Chemistry}, Edited by N.H. March and B. Deb, Academic Publishing, London (1987)


\bibitem{PhysRevA.37.2805}
%  title = {Rayleigh-Ritz variational principle for ensembles of fractionally occupied-states},
Gross, E. K. U. and Oliveira, L. N. and Kohn, W.,
Phys. Rev. A, {\bf 37},
2805, (1988)
doi: 10.1103/PhysRevA.37.2805


\bibitem{rajagopal}
A. K. Rajagopal and F. A. Buot, Phys. Rev. A, {\bf 51}, 1770 (1995)

\bibitem{PhysRevA.87.062501}
E. Pastorczak, N.I. Gidopoulos and K. Pernal, Phys. Rev. A, {\bf 87}, {062501}, (2013); doi:10.1103/PhysRevA.87.062501.






\bibitem{RevModPhys.44.451}
%  title = {Properties and Uses of Natural Orbitals},
Davidson, Ernest R.,
Rev. Mod. Phys.,
{\bf 44},
451, 
(1972); doi: 10.1103/RevModPhys.44.451





\bibitem{PhysRev.180.45}
%  title = {Long-Range Behavior of Hartree-Fock Orbitals},
Handy, Nicholas C. and Marron, Michael T. and Silverstone, Harris J.,
Phys. Rev., {\bf 180},
45,
(1969); 
doi: 10.1103/PhysRev.180.45

\bibitem{PhysRevLett.72.2981}
%  title = {There are no unfilled shells in unrestricted Hartree-Fock theory},
V. Bach, E.H. Lieb, M. Loss, J.P. Solovej, Phys. Rev. Lett., {\bf 72},
2981, (1994)
doi: 10.1103/PhysRevLett.72.2981


\bibitem{koopmans}
T. Koopmans, Physica 1, 104–113 (1934); doi:10.1016/S0031-8914(34)90011-2 

\bibitem{szabo} 
A. Szabo and N.S. Ostlund,
``Modern Quantum Chemistry'', {\it Macmillan Publishing Co. Inc., New York} 
(1996).

\bibitem{ghost}
N. I. Gidopoulos, P. G. Papaconstantinou, and E. K. U. Gross, 
Phys. Rev. Lett. {\bf 88}, 033003 (2002)

\bibitem{k1}
J. Katriel, J. Phys. C: Solid St. Phys., {\bf 13} L375 (1980)

\bibitem{k2}
J. Katriel, Int. J. Quantum Chem. {\bf 23}, 1767 (1983)





\bibitem{us} 
T.W. Hollins, S.J. Clark, K. Refson and N.I. Gidopoulos, 
%Optimized effective potential using the Hylleraas variational method.
{Phys. Rev. B} {\bf 85}, 235126 (2012).

\bibitem{lfx}
T.W. Hollins, S.J. Clark, K. Refson and N.I. Gidopoulos, to appear.


\bibitem{rya}
I.G. Ryabinkin, A.A. Kananenka and V.N. Staroverov,
%Accurate and Efficient Approximation to the Optimized Effective Potential for Exchange
{Phys Rev Lett} \textbf{111}, 013001 (2013).

\bibitem{kohanoff2003density}
J. Kohanoff, N.I. Gidopoulos, {``Density functional theory: basics, new trends and applications''},
in the {\em Handbook of Molecular Physics and Quantum Chemistry}, Edited by Stephen Wilson, Vol. {\bf 2}, Part 5, Chapter 26, pp 532–568, 
John Wiley \& Sons, Ltd, Chichester (2003)



\bibitem{hint}
J.C. Slater {\em Insulators Semiconductors and Metals} Vol. 3 of the series {\em Quantum Theory of Molecules and Solids}, 
McGraw-Hill Book Inc. (1967) [See page 263.]


\end{thebibliography}
\end{document}